\documentclass[11pt]{article}

\usepackage{amssymb}
\usepackage{amsmath}
\usepackage{graphicx}

\newcommand\pubblock{\rightline{\begin{tabular}{l} \pubnumber\\
     \pubdate\\ \hepnumber \end{tabular}}}
\newcommand\pubnumber{UB-ECM-PF-06/24 \\ IRB-TH-3/06}
\newcommand\pubdate{September 2006 }
\newcommand\hepnumber{gr-qc/0609083}

\def\beq{\begin{eqnarray}}    
\def\eeq{\end{eqnarray}}      

\newcommand{\Omo}{\Omega_m^0}

\newcommand{\ORo}{\Omega_{R}^0}

\newcommand{\OLo}{\Omega_{\Lambda}^0}

\newcommand{\OXo}{\Omega_{X}^0}

\newcommand{\OD}{\Omega_{D}}
\newcommand{\ODo}{\Omega_{D}^0}

\newcommand{\rc}{\rho_c}
\newcommand{\rco}{\rho_{c}^0}

\newcommand{\rmr}{\rho_m}

\newcommand{\rD}{\rho_D}
\newcommand{\rDt}{\tilde{\rho}_D}

\newcommand{\rX}{\rho_X}

\newcommand{\wX}{\omega_X}
\newcommand{\wm}{\omega_m}

\newcommand{\amr}{\alpha_m}

\newcommand{\aX}{\alpha_X}
\newcommand{\rL}{\rho_{\CC}}

\newcommand{\rLo}{\rho_{\CC}^0}

\newcommand{\CC}{\Lambda}

\newcommand{\we}{\omega_{e}}

\newcommand{\tOm}{\tilde{\Omega}_m}
\newcommand{\tOmo}{\tilde{\Omega}_m^0}

\newcommand{\tODo}{\tilde{\Omega}_{D}^0}

\newcommand{\tr}{\tilde{r}}

\newcommand{\mysection}[1]{\section{#1}
\renewcommand{\theequation}{\thesection.\arabic{equation}}
\setcounter{equation}{0}}

\begin{document}
\pubblock


 \hyphenation{cos-mo-lo-gi-cal
sig-ni-fi-cant}




\begin{center}
{\large \textsc{Composite dark energy: cosmon models with running
\\ cosmological term and gravitational coupling}} \vskip 2mm

 \vskip 8mm

\textbf{Javier Grande$^{a}$, Joan Sol\`{a}}$^{a,b}$,\
\textbf{Hrvoje
 \v{S}tefan\v{c}i\'{c}}$^{a,c}$ \vskip0.5cm $^{a}$ {High Energy
Physics Group, Dep. ECM, Univ. de Barcelona\\  Av. Diagonal 647,
08028 Barcelona, Catalonia, Spain}

$^{b}$  C.E.R. for Astrophysics, Particle Physics and
Cosmology\,\footnote{Associated with Instituto de Ciencias del
Espacio-CSIC.}

$^{c}$ Theoretical Physics Division, Rudjer Bo\v{s}kovi\'{c}
Institute,\\ P.O.B. 180, HR-10002 Zagreb, Croatia.

E-mails: jgrande@ecm.ub.es, \,sola@ifae.es, \,stefancic@ecm.ub.es

\vskip2mm

\end{center}
\vskip 15mm

\begin{quotation}
\noindent {\large\it \underline{Abstract}}.$\,\,$In the recent
literature on dark energy (DE) model building we have learnt that
cosmologies with variable cosmological parameters can mimic more
traditional DE pictures exclusively based on scalar fields (e.g.
quintessence and phantom). In a previous work we have illustrated
this situation within the context of a renormalization group
running cosmological term, $\Lambda$. Here we analyze the
possibility that both the cosmological term and the gravitational
coupling, $G$, are running parameters within a more general
framework (a variant of the so-called ``$\CC$XCDM models'') in
which the DE fluid can be a mixture of a running $\CC$ and another
dynamical entity $X$ (the ``cosmon'') which may behave
quintessence-like or phantom-like. We compute the effective EOS
parameter, $\we$, of this composite fluid and show that the
$\CC$XCDM can mimic to a large extent the standard $\CC$CDM model
while retaining features hinting at its potential composite
nature (such as the smooth crossing of the cosmological constant
boundary $\we=-1$). We further argue that the $\CC$XCDM models
can cure the cosmological coincidence problem. All in all we
suggest that future experimental studies on precision cosmology
should take seriously the possibility that the DE fluid can be a
composite medium whose dynamical features are partially caused and
renormalized by the quantum running of the cosmological
parameters.
\end{quotation}
\vskip 8mm

\newpage

\vskip 6mm

 \noindent \mysection{Introduction}
 \label{Introduction}


One of the most alluring aspects of modern cosmology is that it has
become an accurately testable phenomenological branch of Physics.
This has granted the field a fairly respectable status of empirical
science, which certainly did not possess some two decades ago.
Undoubtedly the most prominent accomplishment of cosmology to date
has been to provide strong indirect support for the existence of
both dark matter (DM) and dark energy (DE) from independent data
sets derived from the observation of distant
supernovae\,\cite{Supernovae}, the anisotropies of the
CMB\,\cite{WMAP3Y}, the lensing effects on the propagation of light
through weak gravitational fields\,\cite{Lensing}, and the inventory
of cosmic matter from the large scale structures (LSS) of the
Universe\,\cite{LSS}. But in spite of these outstanding
achievements, modern cosmology still fails to fulfill the most
important of its tasks, to wit: unraveling the ultimate physical
nature of the components that build up the mysterious dark side of
the Universe. Ignoring for the moment the puzzling DM sector (which
is recently receiving spectacular empirical support\,\cite{Chandra})
and whose final identity might hopefully be elucidated by directly
detecting some day an exotic form (e.g. supersymmetric) of
elementary particles, the DE component is nonetheless the most
intriguing, unnerving and distressing of all. It turns out to be
constant or slowly varying with time, it does not cluster, it fills
smoothly all corners of our patch of accessible Universe and -- to
pitch even higher our level of perplexity -- it constitutes by far
the dominant form of energy density at the cosmological scale. What
is it? Is it vacuum energy in the form of a cosmological constant?
Is it the value of a slowly evolving homogeneous and isotropic
scalar field? Perhaps a remnant of higher order gravitational terms
in the effective action? Or some hint of a modified gravitational
theory?... We don't know. But we do know at least (as of the early
work of Zeldovich in 1967\,\cite{zeldo}) that the typical size of
the vacuum energy in QFT is exceedingly large to be reconciled with
the characteristic energy densities at the cosmic scale, such as the
cosmological matter-radiation density, the critical density, and of
course also the cosmological energy density ($\rL=\CC/8\pi\,G$)
associated to $\CC$. All of them are of order of $(10^{-3}\,eV)^4$,
far too small compared to the average energy densities in particle
physics, say in the standard model of strong and electroweak
interactions. Unfortunately, moving from QFT to string theory does
not seem to help much, for after the process of compactification
from $11$ dimensions down to $3$ we are left with a vastly complex
``landscape'' consisting of some $10^{1000}$ metastable
(non-supersymmetric) vacua  where to entertain our choice of the
ground state\,\cite{landscape}. All in all it looks very hard to
identify $\rL$ with a vacuum energy density. If we insist on it, we
stumble once again upon the excruciating cosmological constant (CC)
problem whose only known ``technical'' solution at present (though
certainly not a very natural one!) is to postulate the existence of
an \textit{ad hoc} series of extremely fine-tuned cancellations
among the various contributions\,\cite{weinRMP,CCRev}\,\footnote{See
e.g. \,\cite{JHEPCC1} for a summarized presentation.}.

In view of this situation it is perhaps wiser to avoid subscribing
for the moment to a too strong opinion on the ultimate nature of the
$\CC$ term in Einstein's equation. Instead, we may treat it
phenomenologically as a parameter in QFT in curved space-time,
therefore acquiring the status of an effective charge similar to any
running quantity in, say, QED or QCD. If one takes this practical
point of view\,\cite{JHEPCC1}, we may get some guidance from the
powerful renormalization group (RG) methods and use them to extract
testable information on the possible evolution of
$\CC$\,\cite{PeeblesRatra88} from its presently measured value --
see e.g. Ref.\,\cite{RGTypeIa} for a devoted attempt in this
direction. The RG evolution of the CC is not primarily a time
evolution, it stems rather from it being a function $\CC=\CC(\mu)$
of the typical energy scale $\mu$ of the Universe at any given stage
of the cosmic evolution. Following \cite{JHEPCC1} we identify this
energy scale with the expansion rate or Hubble function, $\mu=H$,
which is of course time dependent. Therefore $\CC=\CC(H)$ becomes
also time dependent. The resulting model is a FLRW type cosmological
model with running cosmological parameters. This point of view has
been amply exploited in different ways in
references\,\cite{JHEPCC1}-\cite{Reuter} and constitutes an
alternate candidate to \textit{quintessence}\,\cite{quintessence}.
Both pictures (quintessence and RG cosmology) have in common the
idea of a dynamical DE. This can be useful in that a time-evolving
DE  may help to understand another aspect of the CC problem which is
also rather intriguing, the so-called ``coincidence
problem''\,\cite{CCRev} or the problem of why the presently measured
value of the CC (or DE) is so close to the matter density, i.e. both
being at present of order of $(10^{-3}\,eV)^4$. This problem has
been addressed from different perspectives within modified
quintessence models\,\cite{ModifiedQ}. Remarkably, RG models can
also tackle efficiently the coincidence problem\,\cite{GSS1}, a
point that will be reexamined here too.

One may describe RG cosmological models from the viewpoint of an
effective scalar field parametrization of the DE and compute e.g.
the effective equation of state (EOS) for the running $\CC$. This
was done in detail in \cite{SS1} for a particular case with fixed
gravitational coupling $G$ and with non-conserved CC and matter
densities. In a subsequent work a general algorithm was
established to compute the effective EOS for general cosmological
models with variable cosmological parameters\,\cite{SS2}. A
further generalization (the so-called $\CC$XCDM
models\,\cite{GSS1}) is to admit the possibility that the DE is a
composite medium made out of a running $\CC$ and a dynamical
component $X$ that we called the ``cosmon'' in
\cite{GSS1}\footnote{The name ``cosmon'' was originally introduced
in \,\cite{PSW} to represent a dynamically adjusting DE field.
Here we borrow the name to mean any (effective or fundamental)
dynamical component of the DE (not necessarily quintessence-like)
other than the cosmological term. }. In the present work we
elaborate further along this line but we allow the gravitational
coupling $G$ being variable. In this way we obtain a picture in
which matter and cosmon densities are separately conserved,
whereas $\CC$ evolves together with the gravitational coupling
$G$ according to the Bianchi identity -- thus preserving general
covariance. Apart from $\CC=\CC(H)$ we explicitly obtain the
function $G=G(H)$. The outcome is a variant of the $\CC$XCDM
model in which the change of $\CC$ is compensated for by the
evolution of $G$.


\noindent \mysection {Cosmon models with running $\CC$ and $G$}
\label{CCXCM models}

Consider Einstein's equations with a CC term, $\CC$, and move it
onto the $\textit{r.h.s}$ of the field equations to form the
combined quantity $8\,\pi\,G\,\tilde{T}_{\mu\nu}$. Here
$\tilde{T}_{\mu\nu}$ is the effective energy-momentum tensor
$\tilde{T}_{\mu\nu}= T_{\mu\nu}+g_{\mu\nu}\,\rL$, where
$T_{\mu\nu}$ is the ordinary contribution from matter-radiation,
and $\rL=\CC/(8\pi\,G)$. Next let us impose the Bianchi identity,
$\bigtriangledown^{\mu}\,\left(G\,\tilde{T}_{\mu\nu}\right)=0$.
Upon evaluating it in the FLRW metric one finds
\begin{equation}\label{BD}
\frac{d}{dt}\left[G\,\left(\rmr+\rL\right)\right]+G\,H\,
\alpha_m\,\rmr=0\,,\ \ \ \ \ \alpha_m\equiv 3(1+\omega_m)\,,
\end{equation}
where $\rmr$ and $\rL$ are the matter-radiation and CC densities,
and $\wm=0,1/3\ (\amr=3,4)$ for matter and radiation respectively.
Notice that we contemplate $\dot{\rL}\equiv d\rL/dt\neq 0$ and
moreover we do not drop $G$ from (\ref{BD}) because we also admit
the possibility that $\dot{G}\neq 0$. Let us next assume that the
DE fluid is a composite medium consisting of the CC term and of
another dynamical quantity $X$ (``the cosmon''), with energy
density $\rX$, characterized by an EOS parameter $\wX$. Then it
is easy to see that Eq.\,(\ref{BD}) generalizes into
\begin{equation}\label{generalBD}
\frac{d}{dt}\left[G\,\left(\rmr+\rD\right)\right]+G\,H\,
\left(\alpha_m\,\rmr+\aX\rX\right)=0\,,\ \ \ \ \ \aX\equiv
3(1+\wX)\,,
\end{equation}
where $\rD=\rL+\rX$ is the total DE density. Following
\cite{GSS1} we call the resulting composite DE model the
$\CC$XCDM model. There are, however, different possible
implementations of the $\CC$XCDM model: if we assume that $\rD$
and $\rmr$ are separately conserved, then $G$ must be constant
and we obtain precisely the sort of $\CC$XCDM  framework studied
in detail in Ref.\,\cite{GSS1}, which will be referred hereafter
as type I cosmon models. Alternatively, if we take as conserved
quantities $\rX$ and $\rmr$, then we must have
$\dot{\rho_i}+\,\alpha_i\,\rho_i\,H=0$ for both
$\rho_i=\rmr,\rX$, and in this case (\ref{generalBD}) implies
that $G$ must vary in time in combination with $\rL$ according to
the equation
\begin{equation}\label{BDII}
\dot{G}\,(\rL+\rX+\rmr)+G\,\dot{\rho}_{\CC}=0\,.
\end{equation}
The class of $\CC$XCDM models based on these equations will be
referred to as type II cosmon models and its study is the main aim
of this work. We remark that the type II models constitute a
generalization of the $G$-running cosmological model presented in
Ref.\,\cite{SSS1} where the new ingredient here is the introduction
of the self-conserved cosmon entity $X$ which, as we will see, can
play an important distinctive role. We should stress from the
beginning that the cosmon need not be a fundamental field.  In
particular, $X$ is not necessarily a quintessence scalar
field\,\cite{quintessence}, not even a ``standard'' phantom
field\,\cite{Phantom}. In fact, the barotropic index $\wX$ of $X$
can be above or below $-1$, and moreover the energy density $\rX$
can be positive or negative. Thus e.g. if $\wX<-1$ and $\rX<0$ we
obtain a kind of unusual fluid which has been called ``phantom
matter'' (see Fig.\,1 of \cite{GSS1}). Phantom matter does preserve
-- in contrast to usual phantom energy -- the strong energy
condition, as ordinary matter does. Clearly the cosmon plays the
role of a very general entity ranging from the overall
representation of a mixture of dynamical fields of various sorts, to
the effective behavior of higher order curvature terms in the
effective action. In spite of the phenomenological nature of this
approach at the present stage, we cannot exclude the existence of a
consistent Lagrangian formulation for some of these models. In the
remainder of this paper we focus on the interesting phenomenological
implications of the type II models. Still, it is important to say
that the type II models are motivated by the models with a well
defined Lagrangian formulation such as QFT in curved
space-time\,\cite{JHEPCC1}, to be discussed in the following
paragraphs.

To solve Eq.\,(\ref{BDII}) for $G$ we need some input on the
evolution of $\rL$, otherwise we are led to the trivial solution
$G=const.$ Following \cite{JHEPCC1,RGTypeIa} we shall adopt the RG
inspired evolution equation
\begin{equation}\label{RGEG1b}
\frac{d\rL}{d\ln \mu}=\frac{3\,\nu}{4\,\pi}\,M_P^2\,\mu^2\,,
\end{equation}
where $\nu$ is a free parameter: it essentially provides the
ratio squared of the heavy masses contributing to the
$\beta$-function of $\CC$ versus the Planck mass, $M_P$. We
naturally expect $\nu\ll 1$. As mentioned in the introduction, we
will choose $\mu=H$ as the typical RG scale in
cosmology\,\cite{JHEPCC1}. Assuming a FLRW type of Universe with
flat geometry we finally meet the following set of equations for
the type II cosmon models:
\begin{eqnarray}\label{basicset}
&&H^2=\frac{8\pi\,G}{3}\left(\rmr+\rL+\rX\right)\nonumber\,,\\
&&\rL=C_0+C_1\,H^2\nonumber\,,\\
&&(\rmr+\rL+\rX)\,dG+G\,d\rL=0\,.
\end{eqnarray}
The first one is Friedmann's equation for a spatially flat
Universe, the second equation is the integrated form of
(\ref{RGEG1b}) with the boundary condition $\rL=\rLo$ (the
measured value of the CC density) for $H=H_0$ (Hubble's parameter
at present), hence
\begin{equation}\label{C0C1}
C_0=\rLo-\frac{3\,\nu}{8\pi}M_P^2\,H_0^2\,, \ \ \
C_1=\frac{3\,\nu}{8\pi}\,M_P^2\,;
\end{equation}
and the third equation is an equivalent differential form of
(\ref{BDII}). The basic set (\ref{basicset}) can be analytically
solved to determine $G$ as a function of the scale $\mu=H$, with
the following result:
\begin{equation}\label{GH}
G(H)=\frac{G_0}{1+\nu\,\ln\frac{H^2}{H_0^2}}\,,
\end{equation}
where $G_0=1/M_P^2$. This logarithmic running law for $G$
formally coincides with the one obtained in the framework of
\cite{SSS1}. Of course the time/redshift dependence of $G$ will
be different here because of the cosmon contribution. The form of
(\ref{GH}) suggests that $\nu$ acts also as the $\beta$-function
for the RG running of $G$. To obtain $G=G(z)$ and $\rL=\rL(z)$ as
functions of the cosmological redshift $z$ requires more work on
solving differential equations. We limit ourselves to quote the
final results. The gravitational coupling can only be expressed as
an implicit function of $z$:
\begin{equation}\label{implicit}
\frac{1}{g(z)}-1+\nu\,\ln\left(\frac{1}{g(z)}-\nu\right)=
\nu\,\ln\left[\Omo\,(1+z)^{\amr}+\OXo\,(1+z)^{\aX}+\OLo -\nu\,\right]\,,
\end{equation}
where we have defined $g(z)\equiv{G(z)}/{G_0}$ and
$\Omega_i^0=\rho_i^0/\rc^0$ for the various components, with
$\rc^0$ the critical density at present. Equation (\ref{implicit})
defines implicitly the function $g=g(z)$, and once this is known
the CC density as a function of the redshift is obtained from
\begin{equation}\label{CCz}
\rL(z)=\frac{\rLo+\nu\,\left(\rmr(z)+\rX(z)\right)\,g(z)-\nu\,\rc^0}{1-\nu\,\,g(z)}\,.
\end{equation}
With these equations the type II class of cosmon models is solved.
However, a highly convenient next step to do is to characterize
this model with an effective equation of state (EOS) for its DE
density. This is most useful in order to better compare the
effective behavior of this model with alternative models of the
DE (e.g. quintessence type models).


 \noindent  \mysection{\bf Effective equation of state for the DE in type II cosmon models}
 \label{EffectiveES}

The effective EOS for type I cosmon models was studied in great
detail in \cite{GSS1}. Let us thus concentrate here on type II
models only. The general procedure to obtain the effective EOS
parameter $\we$ for models with variable cosmological parameters
was thoroughly  explained  in \cite{SS1} and \cite{SS2} and we
refer the reader to these and other useful references on this
subject\,\cite{EOSfit}. The formula for $\we$ as a function of
the cosmological redshift is
\begin{equation}\label{effEOS}
\we(z)=-1+\frac{1+z}{3}\,\frac{1}{\rDt(z)}\,\frac{d\rDt(z)}{dz}\,.
\end{equation}
The density $\rDt$ (not to be confused with $\rD$) is the total
DE density in the effective DE picture\,\cite{SS2}. The relation
between $\rDt$ and $\rD$ can be obtained from matching (i.e.
equating) the expansion rates $H$ in the two pictures (in this
case the $\CC$XCDM model and the effective DE picture):
\begin{equation}\label{HH}
G(\rmr+\rD)=G_0\,(\tilde{\rho}_m+\rDt)\,.
\end{equation}
By definition, the effective DE picture\,\cite{SS2} is
characterized by self-conserved DE and matter densities
($\rDt,\tilde\rho_m$) and by a constant gravitational coupling:
$G=G_0$. Generalizing the procedure of \cite{SS2} we can show
that $\rDt(z)$ satisfies the following differential equation
(stemming from the Bianchi identity)
\begin{equation}\label{diffrDt}
\frac{d\rDt}{dz}=\amr\,\frac{\rDt(z)-\xi(z)}{1+z}\,,
\end{equation}
where
\begin{equation}\label{xiz}
\xi(z)=g(z)\,\left(\rL(z)+\frac{\amr-\aX}{\amr}\,\rho_X^0\,(1+z)^{\aX}\right)\,.
\end{equation}
The solution of (\ref{diffrDt}) satisfying $\rDt(0)=\rDt^0$ reads
\begin{equation}\label{rDt}
\rDt(z)=\left(1+z\right)^{\amr}\left[\rDt^0-\amr
\int_0^z\frac{dz'\,\xi(z')}{(1+z')^{(\amr+1)}}\right]\,.
\end{equation}
Again following the methods of \cite{SS2} it is convenient to
rewrite Eq.\,(\ref{HH}) as follows:
\begin{equation}\label{HH2}
\Omo\,f_m(z)\,(1+z)^{\amr}+\OLo\,f_{\CC}(z)+\OXo\,f_X(z)\,(1+z)^{\aX}
=\tOmo\,(1+z)^{\amr}+\tilde{\Omega}_D(z)\,.
\end{equation}
Here $\tilde{\Omega}_D(z)=\rDt(z)/\rco$ with $\rDt(z)$ given by
(\ref{rDt}). The $f_i$ functions above, whatever it be their
form, must satisfy $f_m(0)=f_{\CC}(0)=f_X(0)=1$ in order to
preserve the cosmic sum rule of the $\CC$XCDM models, which reads
(in the flat case)
\begin{equation}\label{sumruletypeII}
\Omo+\ODo=\Omo+\OLo+\OXo=1\,.
\end{equation}
At the same time the parameters $\tOmo$ and $\tODo$ in the
effective DE picture fulfill their own cosmic sum rule
$\tOmo+\tODo=1$. In general the quantities
$\Delta\Omega_m^0\equiv\Omo-\tOmo$,
$\Delta\Omega_D^0\equiv\ODo-\tODo$ will be non-vanishing because
they correspond to different parametrizations of the same
data\,\cite{SS1,SS2}. Of course it only makes sense to
distinguish between $\Omega_i^0$ and $\tilde\Omega_i^0$ when we
fit densities of matter and of DE  e.g. from distant supernovae
data. However, for the radiation component (which is very well
determined by CMB measurements) we set these differences to zero
(see Section \ref{Nucleosynthesis}).

We may now compute ${d\rDt(z)}/{dz}$ from (\ref{HH2}) and insert
the result in (\ref{effEOS}). In the process we use the Bianchi
identity derived in the previous section and the conservation
laws in the $\CC$XCDM models. For the type II cosmon model it is
easy to see that these conservation laws actually entail
$f_m(z)=f_X(z)=g(z)$. And moreover the Bianchi identiy insures
that the following differential expression vanishes identically:
\begin{equation}\label{BI2}
\Omo\,\frac{dg}{dz}\,(1+z)^{\amr}+\OLo\,\frac{df_{\CC}}{dz}+\OXo\,\frac{dg}{dz}\,(1+z)^{\aX}=0\,.
\end{equation}
Using these relations we can obtain, after a straightforward
calculation, the final result. Let us present it in compact form
as follows:
\begin{equation}\label{EOS1}
\we(z)=-1+\frac{\delta(z)}{3\,\tilde{\Omega}_D(z)}\,,
\end{equation}
where
\begin{equation}\label{EOS2}
\delta(z)=\amr\left(g(z)\,\Omo-\tOmo\right)\,(1+z)^{\amr}+\aX\,g(z)\,\OXo\,(1+z)^{\aX}\,.
\end{equation}
As expected, for the special case characterized by $\nu=0$ (no
running $\CC$), $\OXo=0$ (no cosmon) and $\Delta\Omo=0$ (no
difference between the parameters of the two pictures) we obtain
$\delta(z)=0$, for all $z$, and we retrieve the EOS parameter of the
CC:\ $\we=-1$. Recall that the function $g(z)=G(z)/G_0$ has been
obtained implicitly in Eq. (\ref{implicit}) and that it takes the
simple explicit form (\ref{GH}) only when written in terms of $H$.
The latter (log) form suggests that $G(z)$ evolves very little with
$z$ and therefore we expect that the departure of the effective EOS
parameter (\ref{EOS1}) from the CC boundary will be mainly
controlled by the parametrization difference $\Delta\Omega_m^0$ and
by the cosmon contribution -- the last term on the \textit{r.h.s.}
of (\ref{EOS2}). We shall illustrate the behavior of (\ref{EOS1})
with some non-trivial numerical examples in
Section\,\ref{Numerical}. Let us now study the bounds on $\nu$, the
parameter that regulates the evolution of both $\rL(z)$ and $G(z)$.

\noindent  \mysection{\bf Nucleosynthesis bound on $\nu$.
Potential implications for cosmology and astrophysics}
\label{Nucleosynthesis}

As mentioned above, the type II cosmon models furnish a
logarithmic law $G=G(H)$, Eq. (\ref{GH}), which is formally
identical to the one obtained in Ref.\,\cite{SSS1} in which the
cosmon was absent. In the last reference a bound was obtained on
the parameter $\nu$ by considering the experimental limits placed
by nucleosynthesis on the variation of $G$, with the result
$|\nu|\lesssim 10^{-2}$. Here we are going to obtain a more
stringent bound on $\nu$ from the experimental restriction on the
ratio of total DE to matter-radiation densities in the effective
DE picture
$\tr=\tilde{\rho}_D/\tilde{\rho}_m=\tilde{\Omega}_D/\tilde{\Omega}_m$.
This ratio evolves with cosmic time or redshift. Particularly, in
the standard $\CC$CDM model this ratio is given by $\rLo/\rmr$ and
increases without end as time passes by because $\rmr\rightarrow
0$. At present its value is $r_0\equiv\rLo/\rmr^0\simeq 7/3$, i.e.
it is now ``coincidentally'' of order $1$. This is in essence the
cosmic coincidence problem mentioned in the introduction: why is
this ratio of ${\cal O}(1)$ right now? Let us denote the value of
$\tr$ at the nucleosynthesis epoch by
$\tr_N=(\tilde{\rho}_D/\tilde{\rho}_m)_N$. As in previous works
we will use the upper bound $|\tr_N|\lesssim
10\%$\,\cite{GSS1,Ferreira97}. For higher values of $\tr_N$ the
expansion rate at nucleosynthesis would be too large and the
amount of primordial helium synthesized would overshoot the
experimental limits. To convert the upper bound on $|\tr_N|$ into
an upper bound on $|\nu|$ we substitute (\ref{GH}) in
(\ref{implicit}) and solve the resulting equation for $\nu$. The
final (exact) analytical result is
\begin{equation}\label{nu1}
\nu=\frac{\Omo\,(1+z)^{\amr}+\OXo\,(1+z)^{\aX}+\OLo-H^2(z)/H_0^2}
{1+H^2(z)/H_0^2\left[\ln \left(H^2(z)/H_0^2\right)-1\right]}\,.
\end{equation}
Let us use the formula for the expansion rate in the effective DE
picture to eliminate $H^2(z)/H_0^2$ from (\ref{nu1}) in terms of
the ratio $\tr$ defined above:
\begin{equation}\label{HH0}
\frac{H^2(z)}{H_0^2}=\tOm(z)+\tilde{\Omega}_D(z)=\tOmo\,(1+z)^{\amr}(1+\tr)\,.
\end{equation}
At the nucleosynthesis epoch ($z_N\sim 10^9$) we have $\amr=4$
because radiation dominates. Being $z$ very large we can neglect
$\OLo$ in the numerator of (\ref{nu1}). Furthermore, given the
fact that $X$ is assumed to be a DE component
($-1-\delta<\wX<-1/3$, with $\delta>0$ small), we have
$\aX=3(1+\wX)<2$ (hence $\aX<\amr$) and the cosmon contribution
can also be neglected at $z=z_N$. Finally, as mentioned in
Section \ref{EffectiveES}, we set $\Delta\Omo=0$ for radiation.
All in all at the nucleosynthesis epoch Eq.\,(\ref{nu1}) boils
down to
\begin{equation}
\nu\simeq\frac{-\tr_N\,\Omega^0_R(1+z_N)^4}{1+\Omega^0_R(1+z_N)^4(1+\tr_N)\left[
\ln\left(\Omega^0_R(1+z_N)^4(1+\tr_N)\right)-1\right]}
\simeq\frac{-\tr_N}{(1+\tr_N)\,\ln\left[\Omega^0_R\,z_N^4\right]}\,.
\label{nu2}
\end{equation}
Here $\ORo$ is the total radiation density (from photons and
neutrinos) normalized to the critical density at present; $\ORo$
is given by $1.68$ times the standard value of the normalized
photon density $\Omega_{\gamma}^0\simeq 4.6\times 10^{-5}$ (for
$h=0.73$)\,\cite{WMAP3Y}. From these equations the desired
nucleosynthesis bound on $\nu$ ensues immediately:
\begin{equation}\label{nubound}
|\nu|\lesssim 10^{-3}\,,
\end{equation}
which is an order of magnitude more restrictive than the previous
bound obtained in \cite{SSS1}. Since the bound (\ref{nu2})
essentially does not depend on the X (cosmon) component, it also
applies to the models studied in \cite{SSS1}. The fact that
$\nu\ll 1$ is indeed a natural theoretical expectation from the
interpretation of this parameter within effective field theory
(cf. Ref.\,\cite{RGTypeIa} for details). An obvious implication of
this bound on type II cosmon models is that the corresponding
running of $\CC$ is in principle rather hampered as compared to
that in type I models\,\cite{GSS1}. In the latter, the
nucleosynthesis bound affected a combined parameter involving
both the $\CC$ and cosmon dynamics: $\epsilon\equiv
\nu(1+\wX)\lesssim 0.1$. However, for type II cosmon models
nucleosynthesis does not place any bound on the cosmon barotropic
index $\wX$ and in this sense we have more freedom to play with
it. We point out that the ratio $r$ in the $\CC$XCDM picture,
i.e. $r=\rD/\rmr$, is related to $\tr$ as
$r=g(z)^{-1}\,({\tOmo}/{\Omo})\left(1+\tr\right)-1$, where we
have used (\ref{HH}). Let us next consider this expression at the
nucleosynthesis time, where $H(z_N)\equiv H_N$. Setting
$\Delta\ORo=0$, as previously discussed, and using (\ref{GH}) we
find
\begin{equation}\label{rtr}
r_N=\tilde{r}_N+\nu(1+\tilde{r}_N)\,\ln H_N^2/H_0^2\simeq
\tilde{r}_N+\nu(1+\tilde{r_N})\,\ln\left[\Omega^0_R\,z_N^4\right]\,.
\end{equation}
In view of (\ref{nubound}) the second term on the \textit{r.h.s.}
of the above expression remains below $10\%$. Therefore we find
that at the nucleosynthesis time the bound on $r$ lies within the
same order of magnitude as the bound on $\tr$, i.e. $|r_N|<10\%$.
Taking this into account and making the natural assumption that
the value of  $\Delta \Omega^0_m$ is small (big differences
between the two pictures are not expected), we see that in
practice we need not distinguish between $\tr$ and $r$. This is
explicitly corroborated in the numerical analysis performed in
the next section\,\footnote{We point out that the bound on the
relative abundance of DE in the nucleosynthesis epoch
\,\cite{Ferreira97} applies originally to $\tr$ because in that
analysis it is assumed that $G$ is constant. Since, however, this
implies a tight constraint on $\nu$, Eq.\,(\ref{nubound}), it
follows that the bound on $\tr$ essentially applies to $r$ as
well, at least within order of magnitude.}.

The tighter upper limit (\ref{nubound}) on type II models from
nucleosynthesis as compared to type I does not necessarily go to the
detriment of the potentially important role played by $\nu$ when
considering other implications. For example, for sufficiently small
$\nu>0$ (specifically for $\nu\sim 10^{-6}$), the type II cosmon
models can offer an explanation for the flat rotation curves of
galaxies exactly as the RG model of Ref.\,\cite{SSS1}\,\footnote{We
omit details here. See section 7 of Ref.\,\cite{SSS1} for a detailed
discussion of the possible astrophysical implications of a running
$G$ on the flat rotation curves of galaxies. Type II cosmon models
lead to the same kind of picture.}. This is because the type II
models preserve the running logarithmic law (\ref{GH}) wherein at
the astrophysics level $H^{-1}$ is replaced by the radial coordinate
of the galactic system\,\cite{SSS1}. Remarkably, we see that even
for very small (but positive) values of $\nu$ -- far below the
limits placed by cosmology considerations (\ref{nubound}) -- we may
have dramatic implications at the astrophysical scale.

For $\nu<0$, however, these astrophysical implications are not
possible. But in compensation we can get a handle on the
cosmological coincidence problem mentioned in the introduction.
The line of argumentation closely follows the approach of
Ref.\,\cite{GSS1} for type I models. Namely, if the Universe's
evolution has a stopping (or, more properly, a turning) point at
some time in its future, then the ratio $r$ will be bounded from
above for the entire lifetime of the Universe, and therefore our
present time cannot be considered special; that is to say, in a
$\CC$XCDM model $r$ can be of order $1$ not because -- as argued
within the standard $\CC$CDM model -- we just happen to live at a
time near the transition from deceleration into acceleration, but
simply because $r$ can never increase beyond a fixed number.
Moreover, in a large portion of the parameter space $r$ remains
below  ${\cal O}(1-10)$. We refer the reader to \,\cite{GSS1} for
an expanded exposition of this approach to the coincidence
problem, here we limit ourselves to apply the same arguments
within the context of type II models. For these models the
turning point associated to the possible resolution of the
coincidence problem exists only for $\nu<0$. This can be seen
immediately from (\ref{GH}): $H$ will vanish in the future at the
very same point where $G$ also vanishes. However, the stopping
redshift value (call it $z_S$) can be approached with the correct
sign of the gravitational coupling ($G>0$), if and only if
$\nu<0$. For $\nu>0$ we would have $G<0$ before reaching $z_S$.
One can also show that the existence of the turning point can be
realized both for the cosmon barotropic index $\wX$ above or
below $-1$, i.e. for quintessence-like or phantom-like cosmon. To
study the precise conditions for the existence of the turning
point it proves convenient to use Eq.\,(\ref{nu1}) and set
$H(z)=0$ in it. The root $z=z_S$ satisfying the resulting
equation will obviously lie in the matter dominated epoch, so we
can set $\amr=3$ in (\ref{nu1}). Introducing the (continuous)
function
\begin{equation}
f(z)=\Omega^0_m(1+z)^3+\Omega^0_X(1+z)^{3(1+\omega_X)}+(1-\Omega^0_m-\Omega^0_X-\nu)
\end{equation}
we find, equivalently, that $z_S$ is obtained as a root of it:
$f(z_S)=0$. Next we apply Bolzano's theorem for continuous
functions, and to this end we explore the change of sign of $f(z)$
at present ($z=0$) as compared to the remote future ($z=-1$). We
find
\begin{eqnarray}
\label{limit}
\lim_{z\rightarrow -1}f(z)&=&\left\{\begin{array}{ccc}
1-\Omega^0_m-\Omega^0_X-\nu &\textrm{if}& \omega_X>-1\\
{\rm sign}(\Omega^0_X)\cdot\infty &\textrm{if}& \omega_X<-1 \end{array} \right.\\
f(0)&=&1-\nu
\end{eqnarray}
Being $1-\nu$ obviously positive (because $\nu<0$) we can
guarantee stopping point $z_S$ in the future under the following
conditions:
\begin{equation}
\label{stopping1} \Omega^0_X>1-\Omega^0_m-\nu  \ \ \ \ \
\textrm{for}\ \ \ \ \ \omega_X>-1\,,
\end{equation}
\begin{equation}
\label{stopping2}
 \Omega^0_X<0\ \ \ \  \textrm{for} \ \ \ \omega_X<-1\,.
\end{equation}
Notice that $-1<\wX<-1/3$ corresponds to a quintessence-like
cosmon. Given the approximate prior $\Omo=0.3$ (e.g. from
LSS\,\cite{LSS} and supernovae data\,\cite{Supernovae}) and using
the bound (\ref{nubound}) and the sum rule (\ref{sumruletypeII}),
we learn that the corresponding stopping condition
(\ref{stopping1}) is essentially equivalent to require $\OLo<0$.
This should not be considered as a drawback; within our composite
DE framework the generalized sum rule (\ref{sumruletypeII}) makes
allowance for any sign of $\OLo$, provided the total DE density
$\ODo=\OLo+\OXo$ is positive and of the order of the
experimentally measured value $\OD^0\simeq 0.7$. In other words,
this essentially implies $\OXo>0$ and it confirms that the
condition (\ref{stopping1}) is realized by a quintessence-like
cosmon whose density is, therefore, decreasing with time. Being
the running of $\CC$ very mild it is not surprising that the
stopping condition is almost equivalent to $\rho_{\Lambda}^0<0$.
On the other hand, for $\nu<0$ and $-1-\delta<\wX<-1$ we must
necessarily have $\OXo<0$, Eq. (\ref{stopping2}), and in this case
the cosmon behaves like ``phantom matter'' rather than as usual
phantom energy (cf. Fig.\,1 of \cite{GSS1}). Finally, let us
consider the simplest case $\nu=0$ (i.e. cosmon models with fixed
$\CC$ and $G$). In this particular instance the total normalized
DE density is the same as in the type I cosmon models for $\nu=0$
\,\cite{GSS1}:
\begin{equation}\label{ODznu0}
\OD(z)=\OLo+ \OXo\,(1+z)^{3(1+\wX)}\,.
\end{equation}
For $\wX>-1$ (resp. $\wX<-1$) the stopping point exists if
$\OLo<0$ (resp. $\OXo<0$). This reflects the continuity of our
analysis with respect to the variable $\nu$ as this situation
corresponds to the $\nu\rightarrow 0$ limit of the $\nu\neq 0$
case studied above. Remarkably, for any value of $\nu$ the
existence of a turning point for $\wX<-1$ does not correspond to
$X$ behaving as phantom DE but rather as ``phantom
matter''\,\cite{GSS1}. For this reason there is no ``Big Rip''
singularity\,\cite{Phantom} in this kind of models. To summarize,
the existence of a turning point is guaranteed in many
circumstances within the framework of $\CC$XCDM cosmologies, and
it appears to be a very desirable ingredient to help quenching
the severity of the cosmological coincidence problem and at the
same time to eschew future cosmic singularities of the ``Big
Rip'' type. In its absence (namely of a turning point) the
Universe's expansion would be eternal and the ratio $r=\rD/\rmr$
would tend to infinity as in the standard $\CC$CDM model. In the
next section we present some numerical examples illustrating the
various possibilities described above.


 \noindent  \mysection{Numerical analysis}
 \label{Numerical}

The results of the preceding sections are further illustrated in
this section through particular numerical examples. We present
plots of the evolution of the dark energy effective EOS parameter
$\omega_e$, the Hubble parameter $H$ and the ratio of the energy
densities of dark energy and matter components $r$. These plots
depict the interesting phenomena that the type II cosmon models
exhibit in different parameter regimes.
\begin{figure}[t]
\centering
\begin{tabular}{cc}
\resizebox{0.42\textwidth}{!}{\includegraphics{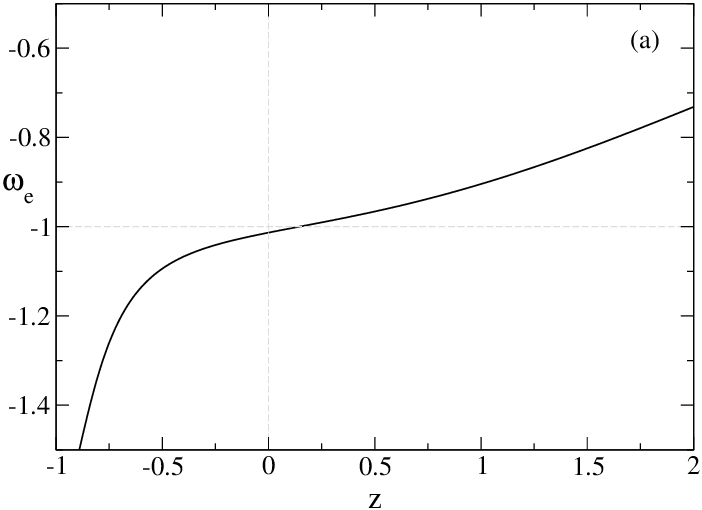}} &
\hspace{0.1cm}\resizebox{0.42\textwidth}{!}{\includegraphics{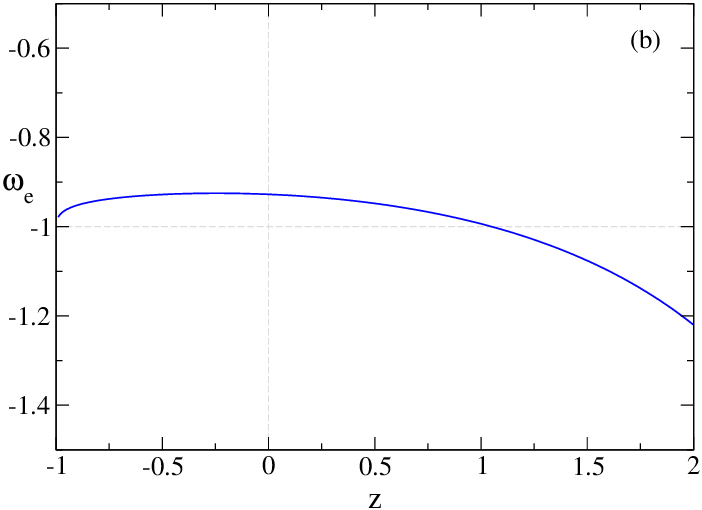}} \\[4.7ex]
\resizebox{0.42\textwidth}{!}{\includegraphics{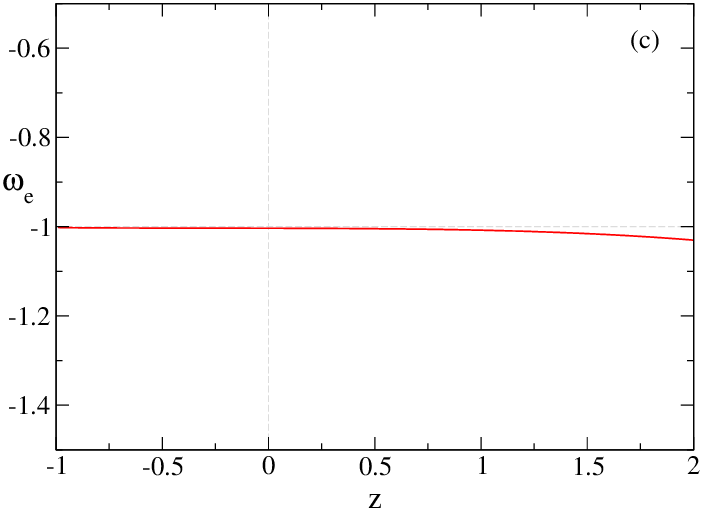}} &
\hspace{0.1cm}\resizebox{0.42\textwidth}{!}{\includegraphics{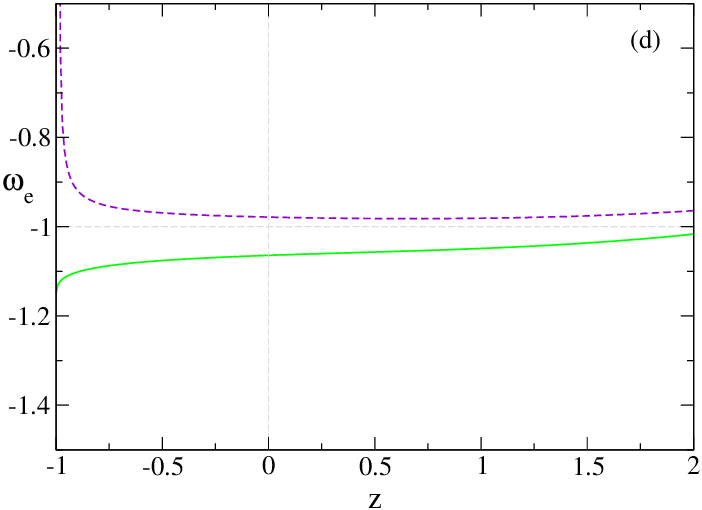}}
\end{tabular}
\caption{Some examples of the behavior of the effective equation of
state function (\ref{EOS1}) for the type II cosmon models: (a)
$\omega_X=-1.65$, $\nu=+0.001$, $\Omega^0_{\Lambda}=0.67$,
$\Delta\Omega^0_m=0.01$. In this case, the EOS presents an effective
transition from quintessence to phantom regime in the recent past;
(b) $\omega_X=-0.85$, $\nu=-0.001$, $\Omega^0_{\Lambda}=0.3$,
$\Delta\Omega^0_m=-0.01$. Here the transition is just the other way
around; (c) $\omega_X=-0.95$, $\nu=0.001$,
$\Omega^0_{\Lambda}=0.75$, $\Delta\Omega^0_m=0$, the EOS mimics the
behavior of a cosmological constant; (d) $\omega_X=-1.15$,
$\nu=-0.001$, $\Delta\Omega^0_m=0$ and $\Omega^0_{\Lambda}=0.4$
(solid) or $\Omega^0_{\Lambda}=0.8$ (dashed). In the last two
examples the EOS parameter remains in the phantom or quintessence
regime for all the redshifts attainable by present and scheduled
supernovae experiments.} \label{plet}
\end{figure}
We first focus on the redshift dependence of the DE effective EOS
parameter $\omega_e$, illustrated in the plots of Fig. \ref{plet}.
Each of these plots demonstrates how type II models can result in
the features of $\omega_e(z)$ which are consistent with the current
observational data. The type II cosmon  models can exhibit a very
interesting phenomenon of the CC boundary crossing, both the
crossing from the quintessence regime to the phantom regime with the
expansion and the transition in the opposite direction. In Fig.
\ref{plet}a we show an example of the realization of the crossing
from the quintessence regime to the phantom regime for the following
values of parameters: $\omega_X=-1.65$, $\nu=+0.001$,
$\Omega^0_{\Lambda}=0.67$ and $\Delta\Omega^0_m=0.01$. An example of
the transition in the opposite direction  for parameter values
$\omega_X=-0.85$, $\nu=-0.001$, $\Omega^0_{\Lambda}=0.3$,
$\Delta\Omega^0_m=-0.01$ is depicted in Fig. \ref{plet}b. The plot
in Fig. \ref{plet}c shows that the $\Lambda$XCDM models may mimic
the behavior of the cosmological constant. The values of the
parameters in this case are $\omega_X=-0.95$, $\nu=0.001$,
$\Omega^0_{\Lambda}=0.75$ and  $\Delta\Omega^0_m=0$. Finally, in the
plot given in Fig. \ref{plet}d we give two examples in which the
type II models exhibit either quintessence-like or phantom-like
features for redshifts amenable to the SNIa observations. An example
of the quintessence-like behavior (the dashed line in Fig.
\ref{plet}d) is obtained for $\omega_X=-1.15$, $\nu=-0.001$,
$\Delta\Omega^0_m=0$ and $\Omega^0_{\Lambda}=0.8$, whereas an
example of the phantom-like behavior (the solid line in Fig.
\ref{plet}d) is realized for $\omega_X=-1.15$, $\nu=-0.001$,
$\Delta\Omega^0_m=0$ and $\Omega^0_{\Lambda}=0.4$. These examples
illustrate the potential of the $\Lambda$XCDM models in explaining
various types of dynamics for $\omega_e(z)$.

\begin{figure}[t]
\begin{center}
\resizebox*{0.6\textwidth}{!}{\includegraphics{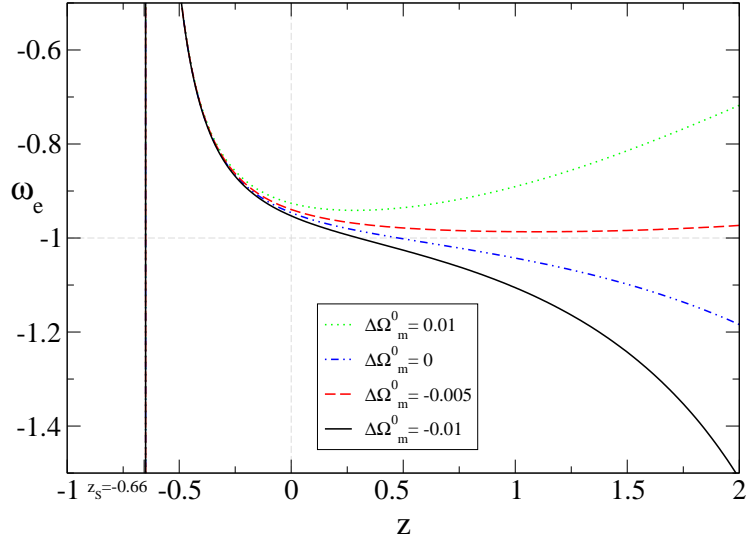}}
\end{center}
\caption{The effective equation of state parameter $\we=\we(z)$
for the case $\omega_X=-1.85$, $\nu=-0.001$,
$\Omega^0_{\Lambda}=0.75$ and different values of
$\Delta\Omega^0_m$. We see that this last parameter crucially
determines the evolution. All the curves present stopping of the
expansion and a ratio between dark energy and matter radiation
bounded from above.} \label{we}
\end{figure}

The dependence of the form of the function $\omega_e(z)$ on the
value of the parameter $\Delta \Omega_m^0$ is given in Fig.
\ref{we} for the following values of the remaining parameters:
$\omega_X=-1.85$, $\nu=-0.001$, $\Omega^0_{\Lambda}=0.75$. From
the figure it is clear that $\omega_e(z)$ function changes
considerably when $\Delta \Omega_m^0$ varies at a percent level,
i.e. it is rather sensitive to the choice of $\Delta \Omega_m^0$.
All curves in this figure are characterized by the stopping of
the expansion of the universe and with $r(z)=\rD(z)/\rmr(z)$
bounded from above.

\begin{figure}
\begin{tabular}{cc}
     \resizebox{0.45\textwidth}{!}{\includegraphics{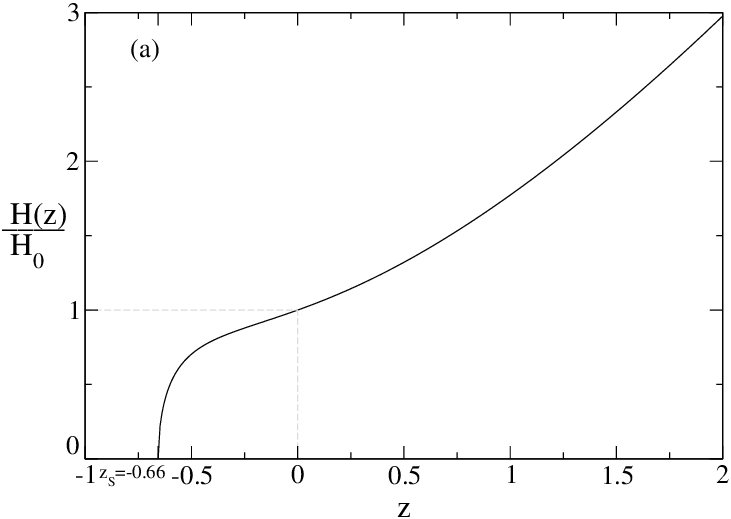}} &
     \hspace{0.1cm}
     \resizebox{0.45\textwidth}{!}{\includegraphics{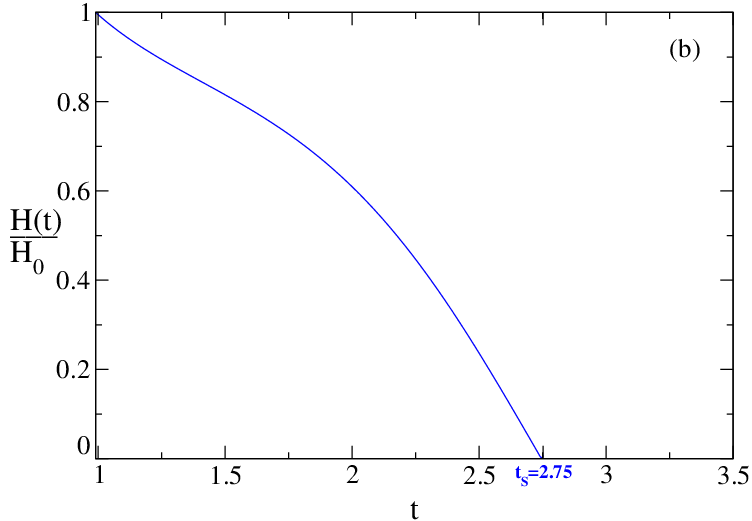}}
   \end{tabular}
\caption{The Hubble function in units of its present value and
depending on (a) redshift or (b) cosmic time (measured in Hubble
time units, $H_0^{-1}$), for the values of the parameters of Fig.
\ref{we} and the specific choice $\Delta\Omega^0_m=-0.005$. The
stopping and subsequent reversal of the expansion takes place after
a time longer than the present age of the Universe,
$t_0=13.7\,Gyr\sim 0.99H^{-1}_0$.}\label{hub}
\end{figure}

\begin{figure}
\begin{tabular}{cc}
     \resizebox{0.45\textwidth}{!}{\includegraphics{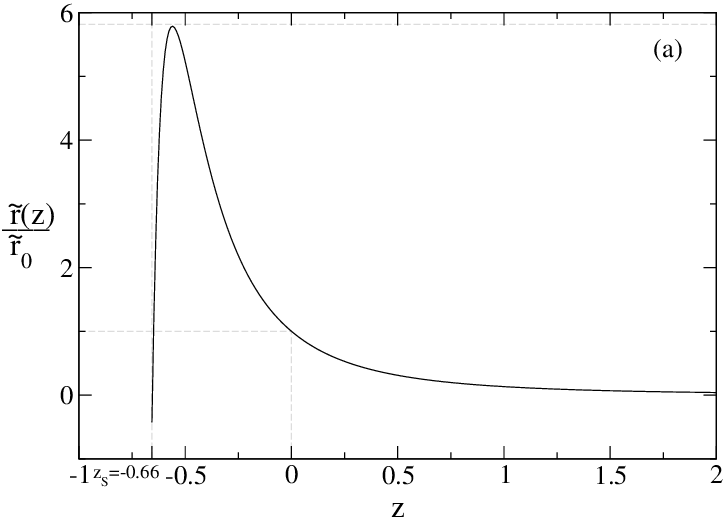}} &
     \hspace{0.1cm}
     \resizebox{0.45\textwidth}{!}{\includegraphics{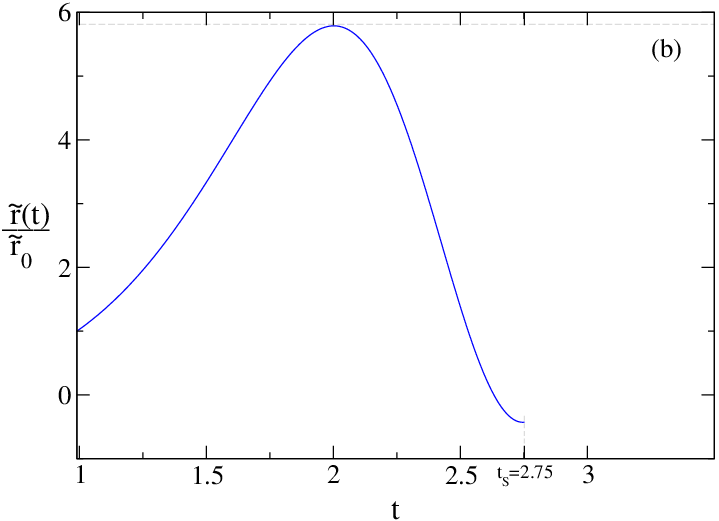}}
   \end{tabular}
\caption{The ratio $\tilde{r}$ between dark energy and matter-radiation in
units of its present value and depending on redshift (a) or
cosmic time (b), for the values of the parameters of Fig.
\ref{we} and the specific choice $\Delta\Omega^0_m=-0.005$. We
see that the ratio presents a maximum that can help to explain or
significantly alleviate the cosmic coincidence problem. The plots for
$r$ result to be nearly indistinguishable from these due to the small
value of $\Delta \Omega^0_m$, and thus are not shown.}\label{rat}
\end{figure}

The discussion presented so far indicates that the $\Lambda$XCDM
models can easily provide the effective dark energy EOS in
accordance with the observational data. This feature is certainly
expected since dark energy contains multiple components and a
correspondingly larger number of parameters. This possible
drawback is, however, more than compensated, by the
characteristics of the type II models which offer a robust
solution to the coincidence problem, as already discussed in
detail in \cite{GSS1} for the type I models. The class of type II
models studied in the present paper also offers the solution of
the coincidence problem. Namely, it is possible to select the
model parameters so that the expansion of the universe stops at
some future moment. An example of such a scenario is depicted in
Fig. \ref{hub}, where the evolution of the Hubble parameter is
given in terms of (a) redshift and (b) cosmic time. The model
parameters used are those from Fig. \ref{we} together with
$\Delta\Omega^0_m=-0.005$. The evolution of the universe of this
type is further characterized by the specific form of the
evolution of the ratio $r$. Namely, $r(z)$ ($r(t)$) is bounded
from above, i.e. it grows with the expansion up to some maximal
value and then starts to decrease, changes sign at some future
moment and attains the value -1 at the stopping point. It is
especially interesting that for a considerable volume in the
model parametric space ($\omega_X$, $\nu$, $\Omega_{\Lambda}^0$)
the ratio $r/r_0$ (where $r_0$ denotes the present value of the
ratio $r$) remains of the order 1 once the universe starts
accelerating. This fact implies that the class of type II models
can provide a solution (or a significant alleviation) of the
coincidence problem. An example of the described dynamics of $r$
in terms of (a) redshift and (b) cosmic time is given in Fig.
\ref{rat} for the same set of parameters as in Fig.
\ref{hub}\,\footnote{For a sufficiently small value of $\Delta
\Omega_m^0$ (which is indeed expected to be so) and $\nu$ values
allowed by (\ref{nubound}), the plots for $r$ and $\tilde{r}$ are
practically indistinguishable. For this reason the ones for $r$
are not included in Fig.\ref{rat}.}. A very similar conclusion
about the evolution of the ratio $r$ was reached within the class
of type I models studied in \cite{GSS1}. The fact that both type I
and type II models provide the solution to the coincidence problem
indicates that this solution is not model dependent and gives
support to a general claim that $\Lambda$XCDM models provide a
robust explanation of the coincidence problem.

Before finishing this section the following observation is in order.
As we have seen, the type II models with flat FLRW metric and a
prior on $\Omega_D^0$ (typically $\Omega_D^0=0.7$, the current value
of the DE density) contain three parameters ($\omega_X$, $\nu$,
$\Omega_{\Lambda}^0$), just as in type I models\,\cite{GSS1}. Of
these three parameters $\nu$ is tightly constrained by the
nucleosynthesis considerations, as described in Section
\ref{Nucleosynthesis}. This parameter space can be compared with
that of other DE models which also try to solve the coincidence
problem. For instance, in interactive quintessence models
(IQE)\,\cite{ModifiedQ} one has a similar number of parameters, to
wit: the EOS parameter of the quintessence field, $\omega_{\phi}$,
the coupling of matter to quintessence (usually in the form of a
source function -- let us call it $Q$ --  which one has to introduce
totally \textit{ad hoc} and depends on at least one parameter), and
finally one or more parameters related to the (assumed) form of the
potential. Moreover, if one wishes to trigger a transition between
quintessence and phantom regimes one usually has to resort to more
complex, e.g. hybrid, structures\,\cite{ModifiedQ} in which at least
one additional scalar field is necessary, though carrying a ``wrong
sign'' kinetic term (i.e. it must be a ghost field). Similar
considerations apply to $k$-essence models\,\cite{kessence}, where
an \textit{ad hoc} non-linear function of the kinetic terms must be
introduced along with additional parameters. In our case we can
modulate that transition through the parameter $\Delta\Omega_m^0$,
as we have seen above. However, even this parameter is not a new
input of the model since it actually belongs to the corresponding
effective EOS picture, not to the $\CC$XCDM model itself. To
summarize, quintessence-like models have in general a similar number
of parameters (if not more) as compared to the $\CC$XCDM, and may
present additional features that one has to ponder carefully. For
example, in IQE models the aforesaid coupling source $Q$ (whose
presence is essential to tackle the coincidence problem) necessarily
entails the non-conservation of matter. This is in contradistinction
to the $\CC$XCDM models, both of type I and II, where matter is
strictly conserved.

 \noindent  \mysection{\bf Conclusions}\quad
\label{Conclusions}
We have studied the possibility that the dark energy (DE) of the
Universe is a composite medium where certain dynamical components
could be entangled with running cosmological parameters. We have
modeled this composite structure by assuming that the DE is a
mixture of a running cosmological term $\CC$ and a dynamical
entity $X$ (the cosmon). The latter is not necessarily an
elementary scalar field, it could rather be the effective
behavior of a multicomponent field system or even the result of
higher order terms in the effective action. In fact the
generality of $X$ is such that its effective barotropic index
$\wX$ can be both quintessence-like ($\wX\gtrsim-1$) and
phantom-like ($\wX\lesssim-1$), and the energy density of $X$ can
be positive or negative ($\rX\lessgtr 0$). The kind of composite
DE model we have studied is a variant of the previously considered
$\CC$XCDM model\,\cite{GSS1}. In all of these models the total DE
density, $\rD$, is the sum of the cosmon density, $\rX$, and the
$\CC$ density, $\rL$. However, in contradistinction
to\,\cite{GSS1}, here the cosmon energy density is conserved (as
the matter density itself), and thus the covariance of Einstein's
equations (expressed by the Bianchi identity) requires that the
running of $\CC$ must be accompanied by a running gravitational
coupling $G$. We have computed the effective equation of state
(EOS) of this composite DE system and found that under suitable
conditions it can mimic to a large extent the behavior of the
standard $\CC$CDM model, but at the same time the fine details
reveal the possibility to observe mild transitions from
quintessence-like into phantom-like behavior, and vice versa. If
observed in the next generation of high precision cosmology
experiments (such as DES, SNAP and PLANCK\,\cite{SNAP}), it would
point at the possible composite structure of the DE. Furthermore,
we have revisited the cosmological coincidence problem within
this model, and found that in a large portion of the parameter
space (similarly to \cite{GSS1}) there is the possibility that the
Universe exhibits a turning point in its evolution, a fact that
would automatically keep the ratio between the DE density to
matter density ($r=\rD/\rmr$) within bounds. Incidentally, it
would also free the cosmic evolution from future singularities
such as the Big Rip\,\cite{Phantom}. We have displayed concrete
examples where the ratio $r$ stays within a few times its present
value $r_0$ for the entire history of the Universe. This feature
seems to be independent of the particular implementation of the
$\CC$XCDM model (whether of type I or of type II as defined in our
analysis) and it suggests that this kind of composite models
could provide a clue to solving, or at least highly mitigating,
the cosmological coincidence problem.

Finally, let us mention that the choice of scale in the RG motivated
approach depends on the symmetries of the underlying space-time
metric.  As an example, in reference \cite{SSS1} it was explored the
possibility to apply the RG approach at the astrophysical level,
which leads the RG scale to develop a dependence on the geometric
coordinates of the system. Furthermore, when we move to the
cosmological context we expect that the $\CC$ term in an
inhomogeneous Universe should develop spatial fluctuations which
could affect the growth of the matter structure and even the CMB
anisotropies. The detailed estimate of the effect of the $\CC$
inhomogeneities on these observables requires an analysis far beyond
the scope of this paper. To get a hint of the kind of investigation
required we refer the reader to the work of\,\cite{FSS1}, which
concentrates on a simpler RG model for $\CC$ in which $G$ is
constant and there is no cosmon. In this case one finds that the RG
running of $\CC$ may have a significant effect on the structure
growth, i.e. that the large scale structure data may provide strong
constraints on the RG parameters. Although the RG effects in
$\CC$XCDM models of type II are not expected to be so important,
owing to the existence of the cosmon component and the severe
nucleosynthesis constraints, the LSS and CMB data are important
tools for placing further constraints on these models. This kind of
considerations will be addressed elsewhere.

To summarize, the $\CC$XCDM models seem to offer a viable
extension of the standard $\CC$CDM model in which some
cosmological problems seem to be better under control. At the same
time they offer a new approach to modeling the DE behavior without
compromising the analysis to a particular field structure of the
theory (e.g. a specific scalar field potential). The dynamics of
these models is entirely determined by the RG behavior of the
cosmological parameters together with the covariant conservation
laws involving the different components of the DE. In our opinion
the $\CC$XCDM models should be considered as serious candidates
for describing the potential DE features observed in the next
generation of experiments.

\vspace{0.1cm}

 \noindent {\bf Acknowledgments.}\
This work has been supported in part by MECYT and FEDER under
project 2004-04582-C02-01, and also by DURSI Generalitat de
Catalunya under 2005SGR00564.  HS thanks the Dep. ECM of the Univ.
of Barcelona for hospitality. He is also supported in part by the
Ministry of Science, Education and Sport of the Republic of Croatia.

\newcommand{\JHEP}[3]{ {JHEP} {#1} (#2)  {#3}}
\newcommand{\NPB}[3]{{\sl Nucl. Phys. } {\bf B#1} (#2)  {#3}}
\newcommand{\NPPS}[3]{{\sl Nucl. Phys. Proc. Supp. } {\bf #1} (#2)  {#3}}
\newcommand{\PRD}[3]{{\sl Phys. Rev. } {\bf D#1} (#2)   {#3}}
\newcommand{\PLB}[3]{{\sl Phys. Lett. } {\bf B#1} (#2)  {#3}}
\newcommand{\EPJ}[3]{{\sl Eur. Phys. J } {\bf C#1} (#2)  {#3}}
\newcommand{\PR}[3]{{\sl Phys. Rep } {\bf #1} (#2)  {#3}}
\newcommand{\RMP}[3]{{\sl Rev. Mod. Phys. } {\bf #1} (#2)  {#3}}
\newcommand{\IJMP}[3]{{\sl Int. J. of Mod. Phys. } {\bf #1} (#2)  {#3}}
\newcommand{\PRL}[3]{{\sl Phys. Rev. Lett. } {\bf #1} (#2) {#3}}
\newcommand{\ZFP}[3]{{\sl Zeitsch. f. Physik } {\bf C#1} (#2)  {#3}}
\newcommand{\MPLA}[3]{{\sl Mod. Phys. Lett. } {\bf A#1} (#2) {#3}}
\newcommand{\CQG}[3]{{\sl Class. Quant. Grav. } {\bf #1} (#2) {#3}}
\newcommand{\JCAP}[3]{{ JCAP} {\bf#1} (#2)  {#3}}
\newcommand{\APJ}[3]{{\sl Astrophys. J. } {\bf #1} (#2)  {#3}}
\newcommand{\AMJ}[3]{{\sl Astronom. J. } {\bf #1} (#2)  {#3}}
\newcommand{\APP}[3]{{\sl Astropart. Phys. } {\bf #1} (#2)  {#3}}
\newcommand{\AAP}[3]{{\sl Astron. Astrophys. } {\bf #1} (#2)  {#3}}
\newcommand{\MNRAS}[3]{{\sl Mon. Not. Roy. Astron. Soc.} {\bf #1} (#2)  {#3}}



\end{document}